\documentclass{amsproc}

\theoremstyle{definition}

\theoremstyle{remark}

\numberwithin{equation}{section}

\begin{document}
\hfill LPT-ENS/01-33
\title{Supergravities: from fields to branes}

\author{Bernard L. JULIA}
\curraddr{Laboratoire de Physique th\' eorique de l'Ecole Normale 
Sup\' erieure, 24 rue Lhomond 75005 Paris France. } 
\email{bernard.julia@lpt.ens.fr}
\thanks{Invited talk presented at the London International Congress
on Mathematical Physics, July 17,-22, 2000. 
The author belongs to UMR 8549 of CNRS.}

\date{November 30, 2000.}


\begin{abstract}
 The quest for unification of particles and fields and for reconciliation 
of Quantum Mechanics and General Relativity has led us to gauge theories,
string theories, supersymmetry and higher-extended objects: membranes...
Our spacetime is quantum mechanical but it admits semiclassical 
descriptions of various ``complementary'' kinds that could be valid 
approximations in various circumstances. 
One of them might be supergravity in 11 
dimensions the largest known interacting theory of a finite number of fields 
with gauged Poincar\' e supersymmetry. 
Its solitons and their dual membranes would be states
in its quantum version called M-theory. We shall review 
the construction of its classical action by deformation of a globally 
supersymmetric free theory and its on-shell superspace formulation. Then
we shall 
focus on the bosonic matter equations of the dimensional reductions   
on tori of dimensions 1 to 8 to exhibit their common self-duality nature.
In the concluding section we shall discuss possible remnants  at the quantum 
level and beyond the massless sector of 
generalised discrete U-dualities. We shall also comment on the variable 
dimension of spacetime descriptions and on the possibility of extending the 
self dual description to spacetime itself and its metric.
\end{abstract}

\maketitle


\section{Extra-dimensions}

If one considers extended objects beyond particles one must choose a number 
of internal dimensions (t,s), ignoring the null case here, as well as the 
signature of  target spacetime (T,S). For a p-brane $(t,s)=(1,p)$. 
Each type (t,s) object is minimally coupled to a ``gauge'' 
potential'': a differential form of degree $s+t$ and is its source. 
Supersymmetric versions of both the brane worldvolume and target superspace
have been considered and allow a geometric description of fermionic theories
with the caveat that some of the most interesting theories
such as 11 dimensional 
supergravity are put on shell by the known such  descriptions.  
In the context of local
field theories one is limited by the spin 2 restriction namely by 
the observation
that under rather general hypotheses one cannot allow interactions of fields 
of spin higher than two and it seems difficult even  to deal with a finite 
number of  spin two fields namely several gravitons. In september 1977 
W. Nahm classified the 
possible superalgebras that are compatible with this restriction in the 
linearised approximation and he found assuming a unique time direction the 
maximal target dimension 
(1,10;32) where the last figure is the number of odd coordinates  
for a Poincar\' e supergravity structure. He also 
recovered the massless spectra of 10d superstring theories ie of their low 
energy field theory limit. Nine months earlier Gliozzi, Scherk and Olive
had identified the sector of the Ramond Neveu-Schwarz  superstring that could 
be supersymmetric 
in target space. The decisive jump from the hadronic mass scale to the 
Planck 
scale for a fundamental realisation of strings as a microscopic theory of 
gravity had taken place in early 1974 see \cite{SS74} and references therein.

In parallel to these developments the construction of  theories invariant under 
rigid supersymmetry   
in 4 dimensions or under their gauged versions went along  and the 
latter, which turn out to be supergravity theories, 
were progressively constructed as  perturbative expansions 
starting with the $N=1$ case
in other words one Lorentz spinor of 4 (odd) supercharges
up to two versions of $N=4$ via the intermediate cases $N=2,3$. The 
$N=8$ case which has precisely the maximal number of supercharges compatible 
with the maximal spin 2 restriction has the same number $8 \times 4 = 32 $ of 
supercharges as the 11 dimensional theory mentioned above. It   
is exactly its toroidal dimensional reduction on a seven dimensional 
internal torus. For more historical references we refer to \cite{J01}. 
The 7 extra-dimensions here play an ephemeral role and make the internal 
symmetry partly geometric as dimensional reduction is compactification followed
by consistent (with the equations of motion) truncation of the theory to the
zeroth Fourier components along the internal dimensions.

To conclude this introduction of extra-dimensions, let us recall that 
they are imposed on us by 
string theory or by supersymmetry, the open question is how precisely 
to extract in a predictive way a low energy 4d approximation. 
The author started working with   5 and 6 dimensions
in 1975 and became really a convert after 
realizing that the Dirac equation lives    virtually in six dimensions; the 
signature $(T,S)$ must be $(3,3)$ 
if one insists on the existence of Majorana-Weyl 
spinors and of course 3 is a fascinating number. One important 
question is to decide what is the relevant dimension for a given problem.
The choice of an appropriate number of 
odd (fermionic) dimensions is also important, even  bosonic problems  may be 
best analysed by  imbedding 
them in superspace to analyse self-duality equations (or BPS
conditions) in their natural setting, we shall comment on these problems in 
section 4.

\section{Deformation of free gauge theories}

The construction of 11 dimensional supergravity action still relies on the 
so-called Noether method \cite{CJS78}. 
It can be seen as a simultaneous  deformation of an infinite 
dimensional abelian gauge algebra equivariant under some global (or rigid)
Lie algebra  and of one invariant of both algebras 
constructed out of a given set of fields, typically a set of  
representations on spacetime or superspace induced
from Lorentz representations. 
This
program has been called the Gupta program and we refer to 
\cite{FF79} for early references. The name Noether comes from the fact that the 
germ of nonlinear deformation of the action is in fact the minimal coupling
to the Noether current 
of the global symmetry one imposes, for instance it is the nonabelian rigid
compact Lie subgroup $G$ of  Yang-Mills theory if one starts with the free 
action for $dim G$ abelian vector potentials. This construction is full of 
ambiguities that  sometimes can be eliminated by field redefinitions and 
which are due for instance to the 
arbitrariness in the Noether current (the so-called improvement terms).
It is also not guaranteed to succeed. This is a typical deformation problem 
and is of a cohomological nature: the obstructions are some cohomology 
classes, the ambiguity is a coboundary  and the germ itself is a cocycle.
Not surprisingly  the relevant cohomology is Lie algebra cohomology, as 
the representations or the (nonlinear) realisations 
including the invariant action can be combined with the 
transformation group itself to form the object one 
is deforming equivariantly under the rigid symmetry to be preserved.

A notorious example of obstruction is the impossibility to add a cosmological 
term to 11d supergravity \cite{BDHS}. It is obvious from the point of 
view of rigid supersymmetry as there is no corresponding de Sitter algebra as
was shown by Nahm already, but starting from 11d Poincar\' e supergravity  
one can show also
that there is no deformation that leads to a local theory with a cosmological 
term. This result may not be so surprising as 
the analogous theory in 4 dimensions,
$N=2$  supergravity, deforms only if one carefully adds a well defined gauge 
coupling to the vector field which has no field theoretic
analogue in  11 dimensions.
 
\subsection{Diffeomorphisms}
In fact the deformation theory of the linearised diffeomorphisms to the true
diffeomorphisms involves some hindsight from differential geometry. One may 
invoke the existence of a deformed Noether identity to extract the 
diffeomorphism transformation law at the first order or use a moving frame 
formalism and appeal to Lorentz covariance  to get the wanted result. In fact
one generally uses a symmetric energy momentum tensor (for instance of a 
matter field) to start the deformation which is not the canonical 
energy-momentum tensor associated to translations as it involves some 
rotational symmetry information as well. It ought
to be possible to clarify this technical difficulty, in any case in practice 
one uses differential  geometry to resum the diffeomorphism deformation
and one concentrates on the fermionic terms and
matter couplings that are the new features here. In the absence of scalars the
result is polynomial and can be found in a few steps. The scalar fields 
appear multiplied by the gravitational coupling constant in a dimensionless 
combination so their non-polynomial contribution is somewhat harder to find, 
this 
was in fact our main motivation for constructing first 
the 11d supergravity, the 
$N=8$ theory in 4d has actually 70 scalar fields and  the 10 dimensional
IIA supergravity has  a scalar dilaton to be dealt with still. Both are 
toroidal compactifications of 11d SUGRA which does not have any scalar field.  

In the general situation one starts from a degenerate, very abelian or very 
solvable structure and deforms it to a generic and rigid one, an infinite 
dimensional version of the deformation of a (pseudo-)Euclidean displacement 
group into a Lorentz or de Sitter group which are indeed simple and rigid. 
Physicists are maybe
more familiar with the converse operation of contraction but this is a 
singular limit whereas the Gupta-Noether procedure
 has a formal (sometimes convergent)
series expansion and belongs to the rich field of deformation theory. 

\subsection{Supergravities}
In 11 dimensions the massless states are the onshell remnants of the 
metric, of a third degree gauge potential with what may seem like an 
abelian gauge invariance and of a spinor valued 1-form that is the gauge 
field of local supersymmetry. 
Altogether one has at the linearised level 44+84 
bosonic states for each momentum and 128 fermionic degrees of freedom.
The free lagrangian is the rather straightforward generalisation of the 
Rarita-Schwinger lagrangian for a spin 3/2 field in 4d plus the bosonic 
quadratic terms to be covariantised under diffeomorphisms.
The rigid nonabelian supersymmetry gives a Noether current 
that leads to a cubic 
coupling, it becomes local by combining an abelian 
supersymmmetry gauge invariance with the global Poincar\' e
supersymmetry
transformation rule. The iterative construction can now start, in fact the 
quadratic fermionic terms in the action are determined together with 
those in the 
transformation law of the fermions by asking for diffeomorphism and Lorentz
covariance with the help of some Clifford algebra identities involving 
actually only the 
subalgebra $Sp(32,{\Bbb R})$  that preserves the Bargmann hermitising matrix 
$\Gamma^0$. 

Higher order terms come from the requirement of supercovariance.
We should stress that the derivation  of an off shell action without 
auxiliary fields (those fields that would be needed to have a true 
representation of supersymmetry) is delicate because the algebra does not 
close off shell. The reason is clear, one has eliminated the (still unknown)
auxiliary fields by using their equations of motion. For bosonic auxiliary 
fields the latter are exchanged with 
the fermionic equations of motion by supersymmetry. Nevertheless the fermionic 
equations must transform without derivatives of the supersymmetry 
parameters in other words involve supercovariant derivatives ie be 
``supercovariant''. This requirement, it turns out,  controls all the quartic 
terms one needs to get an invariant action. 
We refer to the original paper for standard 
factors of 2 to get supercovariant equations out of a 
nonsupercovariant action.  The final check of invariance follows from a rather
formidable Pauli-Fierz... identity somewhat simplified by restriction to the 
symplectic 32x32 matrices. We refer to a recent extension to supergravity in 
first order formalism for more formulas, one result of that paper is that 
beyond 4d the first order formalism (with independent Lorentz connection) is 
not as useful as in 4 dimensions \cite{JS99}. This suggests that the superspace 
formalism should be more involved as well and the restriction to 
a Lorentz connection should  be
relaxed, there are already indications that an abelian gauge group 
should be added but probably the structure is more subtle. 

The superspace formulation of 11d  SUGRA was discovered in 1980 by two different
groups but it was simplified significantly recently \cite{H97}, let us refer 
also to \cite{CGNN00}.

\section{Universal instantons}
There is a famous connection between self-duality equations and existence of 
unbroken  supersymmetry. Let us recall that the so-called BPS condition
started life as a solvable limit of dyon solutions where the similarity 
between the adjoint  Higgs field and a spacelike extra component of the 
Yang-Mill potential becomes exact, there was no fermion in the picture. It is 
the stationary version of the famous instantonic self-duality equation of 
pure Euclidean Yang-Mills theory. Subsequently
and case by case a suitable supersymmetric extension of each theory
admitting ``self-dual'' solutions was 
always constructed in which  
a Killing spinor ie a covariantly constant spinor can be interpreted as 
an  unbroken supersymmetry of the bosonic 
background which implies the saturation of the Bogomol'ny bound. The first 
analysis of this phenomenon was given in \cite{OW78} in the case of rigid 
supersymmetry. So a
bosonic self-duality equation becomes the condition of preservation of some 
supersymmetry and stability can be reinterpreted as the property of the 
supersymmetry algebra that some bosonic generators are sums of squares of 
fermionic ones. 
In the bosonic case the converse of dimensional reduction has 
been first coined group disintegration and then oxidation, we are advocating 
now a superoxidation mechanism. 

In a way the next sections address the opposite 
problem we are going to show that all the bosonic matter equations of toroidally
compactified 11d SUGRA can be rewritten as self-duality equations of a 
generalised but universal type once one doubles the field content. It is a 
standard procedure in the analysis of differential systems to introduce 
auxiliary variables to render the system first order. The nontrivial 
observation maybe is now
that our rather intricate systems are always defined by a finite 
dimensional superalgebra (ie ${\Bbb Z}_2$-graded Lie algebra) and have a 
universal form. The occurrence of fermionic symmetries is surprising 
for bosonic equations but can be understood from the odd character of odd 
degree gauge potentials like the three form of 11d SUGRA
\cite{CJLP98}.

We shall call self-duality equation any equation relating some curvatures 
$\mathcal{F}$ and
 of the form 
\begin{equation}
\mathcal{F} = * S \mathcal{F},
\end{equation}
where $S$ is an operator of square plus or minus one that compensates for the 
same property of the Hodge duality, but more fundamentally $S$ exchanges the 
generators of the superalgebra associated to gauge potentials and those 
associated to their duals.

\subsection{Middle degree}

The prototype examples are of course the 4d
Maxwell equations written in terms of 
electric and magnetic potentials with dual field strengths. Similarly in  
2d the principal sigma model or more generally the
symmetric space sigma models can be rewritten in the above form, at least for 
the propagating degrees of freedom. We recall that the typical structure is that
of a coset space $KG\backslash G$ where $KG$ is the maximal compact
subgroup of $G$. There are two descriptions, first 
the gauge fixed one where one 
chooses a representative of each coset but the better one restores the 
$KG$ gauge 
invariance and allows the symmetry under $G$ to become manifest. In the latter 
case however the self-duality (in 2d at this stage) involves the components of 
the field strength orthogonal to $KG$ only
\begin{equation}
\mathcal{F} = (dg.g^{-1})^{\perp}.  
\end{equation}
 In fact a harmonic scalar function  and its conjugate form a first order self 
dual pair and one can restore the $SL(2,{\Bbb R})$ invariance subgroup of the 2d
conformal group by the same trick. 

The main example that led to our discovery of the general structure is the 
case of the 28 vector potentials of 4d $N=8$ SUGRA that cannot form a 
representation of the duality symmetry group $G=E_7(7)$ unless one 
combines them with the 28 (Hodge) duals. The scalar fields in that theory obey 
the equations of the sigma model $KG\backslash G$ again  and the self-duality 
equation for the vectors reads in that case
\begin{equation}
g. \mathcal{F} = * S g. \mathcal{F},  
\end{equation}
where $g$ stands for the 56 dimensional matrix representation of $ G$ and 
S has to be an invariant operator for $KG=SU(8)$ \cite{CJ79}. This 
structure has 
been extended to the compactifications of 11d SUGRA on a 3-torus and on 
a 5-torus in \cite{CJLP97} for the field strengths of degree half that of the 
spacetime volume form. 

\subsection{Self-duality for all forms}
From there it was natural to try an extension to all fields, and we succeeded 
for all bosonic forms leaving aside for the time being the graviton and the 
fermions. We expect the latter to transform only under the compact subgroup $KG$
and under the Lorentz (spin) group. We are now  going to exhibit a vast 
generalisation of $G$ or at least of its Borel subgroup. This is quite typical 
of broken symmetries in polynomial situations in which the components of some 
group element $g$ appear also polynomially in its inverse $g^{-1}$ which occurs 
also as we have seen in the equations. The way to permit this is of course 
nilpotence
and  this is why the coset spaces appear usually in their Iwasawa 
parametrisation. One must restore the local $KG$ invariance to have simple
formulas for the fermionic couplings and for the full action of $G$. 
We refer to \cite{CJLP98} for the compactified cases but we shall 
illustrate our general structure in the 11 dimensional case; the 
4-form field strength has a dual that has a non abelian piece. A compact way to 
encode the equation of motion  and the 
Bianchi identity is to define a supergroup 
element and its field strength or curvature by
\begin{equation}
\mathcal{E} = exp(A_3 T) exp (A'_6 T') 
\end{equation}
\begin{equation}
\mathcal{F} = d\mathcal{E} \mathcal{E} ^{-1}. 
\end{equation}
This is a generalised sigma model 
structure, one pair of generators for each form and its 
dual,  a theory is then specified by the choice of a  supergroup law. The 
action of the involution  $S$ is simply the exchange of T and T'.
11d SUGRA is defined by the superalgebra
\begin{equation}
\{T,T\}_+ = T'.
\end{equation}

\section{Conclusion}
\subsection{Discrete symmetries: arithmetic groups}
It is now increasingly plausible that the internal symmetries of the massless 
sectors of the various compactifications of string theories are continuous 
versions of discrete groups of symmetries of the full theories. This was argued 
by A. Sen, P. Townsend, C. Hull, E. Witten and more recently by M. Green and 
his collaborators. Assuming these duality symmetries does allow us to control 
nonperturbative effects and sometimes even to resum perturbation series when 
the space of ``modular forms'' is  small enough. Striking dualities relate
different parametrisations, for instance the size of the periodic eleventh 
dimension of maximal Supergravity emerges as the IIA string coupling constant 
in 10 dimensions and the Planck lengths are related as usual.
\subsection{Doubled spacetime}
The most striking property of our self duality results in my opinion is that
gravitation that is somehow spectator in 11 dimensions enters the game and 
fuses completely with matter forms as one descends to 3 spacetime dimensions 
by toroidal compactification.
I have been emphasizing this repeatedly  and I consider this as a challenge 
for my lifetime. I proposed a provocative picture at the conference where the 
main missing link was precisely a kind of ``Doubled spacetime'' that would 
implement the selfduality on spacetime itself and on its geometry.
I reproduce a diplomatic version of the picture in fig. 1.
 
As a conclusion I may just mention that since the conference
the Coxeter groups of hyperbolic 
Kac-Moody algebras exhibited in \cite{J82} have been discovered to control 
chaos and not only symmetries of the homogeneous reduction of supergravities to 
one dimension of time. As for the main problem namely the extension of our
self-dual formalism to the gravitational sector, little has been established 
beyond the linear level yet. It is important to keep in mind that the full 
diffeomorphism groups should appear and not only their linear subgroups which
have already led to a rather precise systematics for the extension and 
overextension of Dynkin diagrams. It is not the place here to list speculative 
papers or incomplete analyses of facts that indeed make me believe there is a 
big surprise waiting for us around the corner. 


\begin{figure}[tb]
\caption{The 2000 version of a theoretical cathedral}
\label{firstfig}
\[ \begin{array}{cccccc}
   &  & M----- & Theory &  &   \\
& D-Brane &\mathcal{D}-space? & Sugras& Guts &   \\
& Anomalies& Sustrings& Std.Model& Solitons &\\
& Wick& Strings&  Fermi& Higgs& \\
Einstein& Relativities& & &Qed& Dirac \\
Poincar\acute{e}& Lorentz&  & & Maxwell& Schroedinger\\
Newton& Gravity&  & & Optics& Galileo\\
   &  & GEOMETRY/ & /QUANTA &  &   .

\end{array} \]
\end{figure}


\bibliographystyle{amsalpha}

\end{document}